\documentclass{llncs}
\usepackage[pdftex]{graphicx,color}
\usepackage{tikz}
\usepackage{pgf}

\usepackage{makeidx}  

\newcommand{\ie}{\emph{i.e.,} }
\newcommand{\eg}{\emph{e.g.,} }
\begin{document}
%
%

\title{Parallel Local Search: Experiments with a PGAS-based programming model}

\author{Rui Machado\inst{1,2} \and Salvador Abreu\inst{2} \and Daniel Diaz\inst{3}}

\institute{
Fraunhofer ITWM,  Kaiserslautern, Germany \\
\email{rui.machado@itwm.fhg.de} \\
\and
Universidade de \'Evora and CENTRIA, Portugal \\
\email{spa@di.uevora.pt} \\
\and
University of Paris 1-Sorbonne, France \\
\email{Daniel.Diaz@univ-paris1.fr}} 
\maketitle
\makeatletter
\makeatother

\begin{abstract}
  Local search is a successful approach for solving combinatorial
  optimization and constraint satisfaction problems. With the
  progressing move toward multi and many-core systems, GPUs and the
  quest for Exascale systems, parallelism has become mainstream as the
  number of cores continues to increase. New programming models are
  required and need to be better understood as well as data structures
  and algorithms.  Such is the case for local search algorithms when
  run on hundreds or thousands of processing units.  In this paper, we
  discuss some experiments we have been doing with Adaptive Search and
  present a new parallel version of it based on GPI, a recent API and
  programming model for the development of scalable parallel
  applications.  Our experiments on different problems show
  interesting speedups and, more importantly, a deeper interpretation
  of the parallelization of Local Search methods.

\keywords{Parallel Local Search, GPI, Adaptive Search, Constraint Programming}
\end{abstract}

\section{Introduction}

Systematic and complete search algorithms impose a limitation on the
problem size they are able to solve due to the exponential increase in
processing time and memory requirements. For this reason,
heuristics-based search algorithms are used (and necessary) for larger
problem sizes. Instead of exploring the complete search space,
heuristics are used to guide the search to portions of the search
space where solutions might be found. Local Search and Meta-heuristics
are an interesting paradigm for combinatorial search and have been
shown very effective for solving real-life problems
\cite{metaheuristics,handbook-approx}. But despite the effectiveness
of local search methods, for really large problem instances, the
running time required might still be substantial. One way to cope with
this problem is by introducing parallelism.

The current trend we are facing is an inevitable paradigm shift
towards multicore technologies where parallelism is now
omnipresent. In recent systems parallelism spreads over several
systems levels and heterogeneity is growing on the node as well as on
the chip level. Data must be maintained across a hierarchy of memory
levels and most applications and algorithms are not yet ready to take
full advantage of available capabilities. There is a demand for
programming models with a flexible threads model and asynchronous
communication to cope with this gap.

PGAS (Partitioned Global Address Space) programming models have been
discussed as an alternative to MPI~\cite{MPI-2.2} for some time. The
PGAS approach offers the developer an abstract shared address space
which simplifies the programming task and at the same time facilitates
data-locality, thread-based programming and asynchronous
communication. GPI is a PGAS API that follows this philosophy and
delivers the full performance of RDMA-enabled\footnote{RDMA - Remote
  Direct Memory Access.}  networks directly to the application without
interrupting the CPU.

In this paper we aim at bringing together both the need for
parallelism to solve large problem instances with Local Search and its
availability in current systems. We implemented a new parallel version
of the Adaptive Search algorithm based on GPI that goes beyond the
simple independent multiple-walk. Our new design shows interesting
speedup gains on benchmarks with scalability problems and more
importantly, a deeper interpretation on the parallelization of
Adaptive Search in particular and Local Search methods in general,
based on some characteristics of the benchmarks.

The rest of the paper is organized as follows: in
section~\ref{sec:gpi} we present GPI and its programming model,
hightlighting some its major features. Section~\ref{sec:par_ls}
provides some background on the Adaptive Search algorithm and section
\ref{sec:par_as} focuses on its parallelization. In
section~\ref{sec:as_gpi}, we detail our parallelization strategy based
on GPI and in section~\ref{sec:exp_res} we show the obtained results
and compare it to the previous
implementation. Section~\ref{sec:discussion} examines and interprets
our experimental findings, correlating them with the characteristics
of the problems. Finally, section~\ref{sec:conclusion} presents a
short conclusion and perspectives of future work.

\section{GPI}
\label{sec:gpi}
\begin{figure}
\centerline{\fbox{{\includegraphics[scale=0.8]{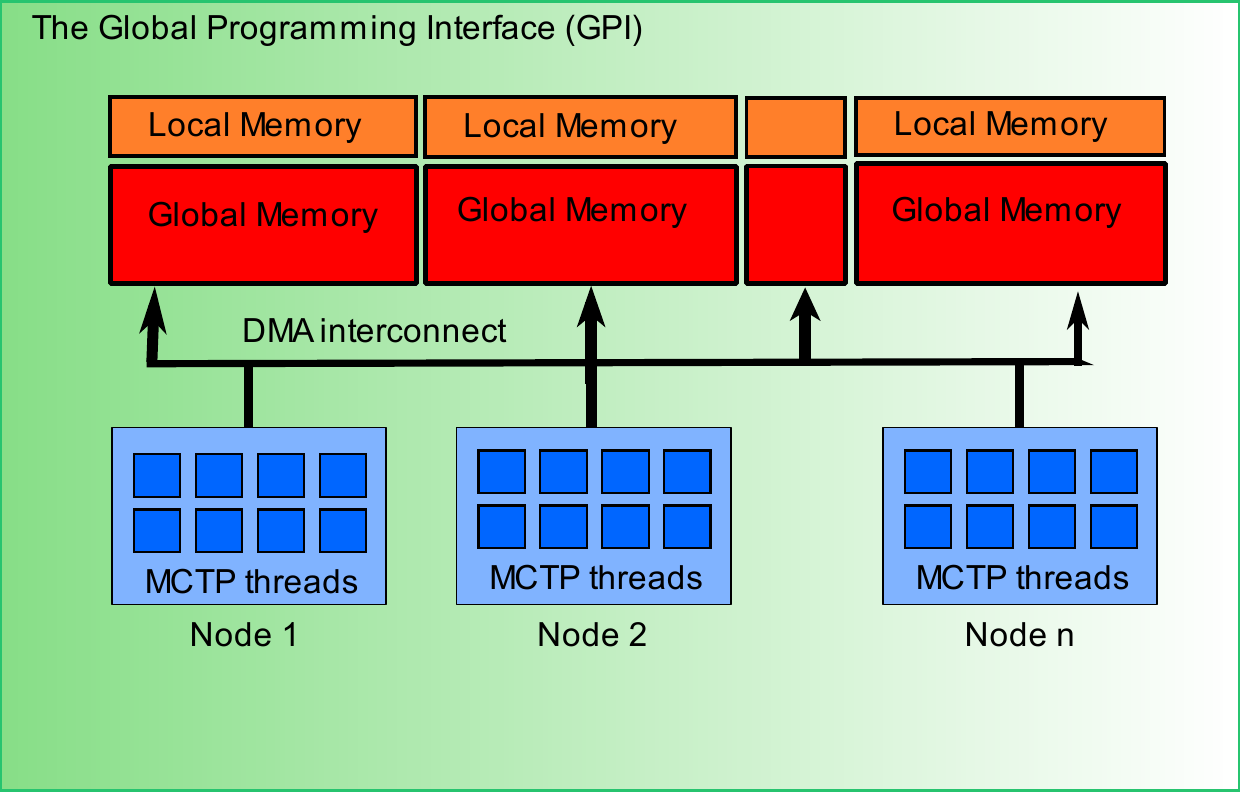}}}}
\caption{GPI}\label{fig:gpi_arch}
\end{figure}

GPI (Global address space Programming Interface) is a PGAS API for
parallel applications running on clusters. The thin communication
layer delivers the full performance of RDMA-enabled networks directly
to the application without interrupting the CPU.
Fig.~\ref{fig:gpi_arch} depicts the architecture of GPI.

The local memory is the internal memory available only to the node and
allocated through typical allocators (e.g. malloc). This memory cannot
be accessed by other nodes. The global memory is the partitioned
global shared memory available to other nodes and where data shared by
all nodes should be placed. The DMA interconnect connects all nodes
and is through this interconnect that GPI operations are issued.  At
the node level, the Manycore Threading Package (MCTP) is used to take
advantage of all cores present on the system and make use of the GPI
functionality and global memory.  The MCTP was developed to help
programmers take better advantage of new architectures and ease the
development of multi-threaded applications.  The MCTP is a threading
package based on thread pools that abstracts the native threads of the
platform.

GPI is constituted by a pair of components: the GPI daemon and the GPI
library. The GPI daemon runs on all nodes of the cluster, waiting for
requests to start applications and the library holds the functionality
available for a program to use: read/write global data, passive
communication, global atomic counters, collective operations.  The two
components are described in more detail in our previous
contribution~\cite{fvm}.\footnote{GPI was previously known as
  Fraunhofer Virtual Machine (FVM).}

The GPI core functionality can be summarized as follows:

\begin{itemize}
\item read and write global data
\item passive communication
\item send and receive messages
\item commands
\item global atomic counters and spinlocks
\item barriers
\item collective operations
\end{itemize}

\noindent
In the context of this work, the most important functionality is the
read/write of global data. 

Two operations exist to read and write from global memory independent
of whether it is a local or remote location. One important point is
that those operations are one-sided that is, only the peer that issues
such operation takes part in it. This is different from a two-sided
scheme (message passing) where the peer that sends (\emph{sender}) has
a corresponding peer (\emph{receiver}) that needs to issue a receive
operation. Moreover, this functionality is non-blocking and completely
off-loaded to the interconnect, allowing the program to continue its
execution and hence take better advantage of CPU cycles. The data
movement does not require any intermediate buffers and protocols to
maintain those buffers. If the application needs to make sure the data
was transferred (read or write), it needs to call a wait operation
that blocks until the transfer is finished and asserting that the data
is usable.

\section{Adaptive Search}
\label{sec:par_ls}

Local Search is based on the simple idea of ``searching'' by
iteratively moving from one solution to one of its
\emph{neighbours}. The neighborhood of a solution is the set of
solutions that can be obtained by applying a \emph{move}. A
\emph{move} is a local change (hence the name Local Search).

The mechanism used to select a neighbour and thus the definition of
what constitutes a neighbourhood is the main issue that differentiates
between different local search methods. In general, it is problem
dependent and is related to the definition of the \emph{objective
  function}.

The Adaptive Search method~\cite{codognet2001yet} is one of many
different local search methods and has proved to be very efficient in
the types of problems where it was tested. It is a generic,
domain-independent constraint-based local search method.

This meta-heuristic takes advantage of the structure of the problem in
terms of constraints and variables and can guide the search more
precisely than a single global cost function to optimize, such as for
instance the number of violated constraints.  The algorithm also uses
an short-term adaptive memory in the spirit of Tabu Search in order to
prevent stagnation in local minima and loops.  This method is generic,
can be applied to a large class of constraints (e.g. linear and
non-linear arithmetic constraints, symbolic constraints, etc) and
naturally copes with over-constrained problems. 

The input of the method is a problem in CSP format, that is, a set of
variables with their (finite) domains of possible values and a set of
constraints over these variables.  For each constraint, an ``error
function'' needs to be defined; it will give, for each tuple of
variable values, an indication of how much the constraint is violated.
For instance, the error function associated with an arithmetic
constraint $|X - Y| < c$, for a given constant $c \geq 0$, can be
$max(0, |X-Y|-c)$. 

Adaptive Search relies on iterative repair, based on variable and
constraint error information, seeking to reduce the error on the worst
variable so far. The basic idea is to compute the error function for
each constraint, then combine for each variable the errors of all
constraints in which it appears, thereby projecting constraint errors
onto the relevant variables. Finally, the variable with the highest
error will be taken and its value will be modified. In this second
step, the well known min-conflict heuristic is used to select the
value in the variable domain which is the most promising, that is, the
value for which the total error in the next configuration is minimal.
In order to prevent being trapped in local minima, the Adaptive Search
method also includes a short-term memory mechanism to store
variables to avoid (variables can be marked Tabu and ``frozen''
for a number of iterations).  It also integrates reset transitions to
escape stagnation around local minima.  A (partial) reset consists in
assigning fresh random values to some variables (also randomly
chosen). A reset is guided by the number of variables being marked
Tabu. As in any local search method, it is also possible to restart
from scratch when the number of iterations reaches a given limit.

\section{Parallel Adaptive Search}
\label{sec:par_as}

When parallelizing an algorithm one aims at identifying hotspots and
sources of parallelism. As with most of meta-heuristics, in Adaptive
Search these sources of parallelism are essentially: (1) the inner
loop of the algorithm \ie computing and combining the errors of
variables and selecting the variable with highest error and (2) the
search space of the problem.

\noindent
The problem with exploiting the inner loop of the algorithm is its
granularity: it is too fine-grained. The overhead associated with
synchronization and dispatching of tasks comes at a too high cost.

The other main source of parallelism is the search space (domain) of
the problem itself. Theoretically, this domain could be decomposed in
several disjunct partitions to be explored in parallel and without
dependencies. However, in practice, several issues arise with this.
Each partition is in general still too large for a sequential
execution and more importantly, not the whole search space is equally
valid and the exploration should avoid areas of it that lead to poor
solutions. Moreover, it is hard and expensive to control and maintain
the search conducted in the different partitions since a Local Search
algorithm only has a local view of the search space. One example is
the class of problems that have the best solutions clustered in a
certain 'zone' of the search space. In this case, the algorithm should
converge to that zone but in case of parallel execution avoid too much
redundant work.

The Adaptive Search method has already been subject to some research
on its parallel behaviour. Previous work on parallel implementations
of the Adaptive Search algorithm have mostly focused on independent
multiple-walks, requiring no commmunication neither shared memory
between processing units.

In~\cite{DBLP:journals/concurrency/DiazAC12}, the authors present a
parallel implementation of the Adaptive Search algorithm for the
Cell/BE, a heterogenous multicore architecture. The system includes 16
processors (the SPEs) where each one starts with a different random
initial solution. The PPE acts as the master processor, waiting for the
message of a found solution. For such number of processing units, the
results were very promising, achieving for some problems linear
speed-up.

Further work with Parallel Adaptive Search continued to follow the
same approach with no communication between workers but more
interestingly, concentrating on cluster systems with a larger number
of cores.

In~\cite{DBLP:conf/evoW/CaniouCDA11}, the authors experiment and
investigate the performance of a multiple independent-walk on a system
with up to 256 cores. The parallelization was done with MPI and
involves the introduction of a ``communication step'' which tests if
termination was detected (a solution was found) and terminates the
execution properly.

The presented performance results are relatively modest in terms of
parallel efficiency and still far away for the ideal speed-up which
contrasts with the results obtained at a smaller scale (ie. up to 16
cores) in previous work. This points out the need for better
alternative strategies in order to better exploit large-scale
parallelism.

Since that the independent multiple-walk approach still leaves space
for improvement in terms of parallel efficiency and scalability for
some problems, new ways to take take full advantage of parallel
systems must be found. 

In \cite{caniou11}, the authors experiment with more complex
strategies, where processes exchange messages resembling
branch-and-bound methods where the bound is exchanged between all
participants. In their work, two alternatives are attempted:
exchanging the cost of the current solution of each process and
the current cost plus the number of iterations needed to achieve that
cost. Unfortunately, both approaches do not achieve better results
than an independent multiple-walk.

\section{Adaptive Search with GPI}
\label{sec:as_gpi}
Previous work with parallel Adaptive Search provides some groundwork
to build upon and has showed that some benchmarks exhibit scalability
problems when run on a large number of cores.

\noindent
GPI seems, \emph{\`a priori}, an interesting match to the problem of
parallelization. Local search methods work with local information,
trying to progress and converge to solutions in a global search space,
requiring low global information. However, in a parallel setting,
communication and cooperation are crucial and in this case, required
to overcome the low parallel efficiency in some problems. The
communication with GPI is based on one-sided primitives that might
benefit the local view on a global search space, allowing threads to
cooperate asynchronously. Moreover, communication is very efficient as
GPI exploits the full performance of the interconnect. Hence, we
continue to explore ways to further improve the parallelization of the
Adaptive Search algorithm, exploiting GPI and its programming model,
with the objective of getting some further benefits. But more
importantly, to find mechanisms, concepts or limitations that are
general.

In general, we can define the following objectives:
\begin{itemize}
\item further investigate and understand the behavior of parallel
  Adaptive Search on different problems.
\item investigate the possibilites given by GPI and devise more
  complex mechanisms for the parallel execution of Adaptive Search,
  improving its performance
\item identify the, possibly new, problems generated by the previous point.
\end{itemize}

\noindent

The new parallel version of Adaptive Search based on GPI includes two
variants which we name TDO (Termination Detection Only) and PoC
(Propagation of Configuration). 

The TDO variant implements the simple independent multiple-walk and
serves mostly has our basis for comparison. First, with the existing
MPI version, making sure that the implementation is correct and the
performance is as expected. Second, to allow us to measure the
improvement (if present) obtained with the more complex PoC
variant. The PoC variant introduces more communication and sharing
between working threads, by means of GPI primitives and threaded
model.

The next sections present the two different variants in more detail.

\subsection{Termination Detection Only}

The variant with Termination Detection Only (TDO) is rather
straightforward and implements the idea of an independent
multiple-walk: all available cores execute the sequential version of
the Adaptive Search algorithm.

We name this variant as Termination Detection Only since it subsumes
itself to a termination detection problem \ie detecting the
termination of a distributed computation. Termination Detection is
itself a subject of much research and several algorithms have been and
continue to be
proposed(~\cite{DBLP:journals/ipl/DijkstraFG83,mattern87,DBLP:conf/ppopp/SaraswatKKGK11}).

In the case of the Parallel Adaptive Search method, we are interested
in detecting termination as soon as one of the participating threads
has found a solution, instead of waiting for all threads to finish as
some of them can potentially require too many steps in order to find a
solution (it is enough to be trapped in a 'zone' of the search space
with no possible solutions).

The implementation of this variant is simple as it only involves the
implementation of a mechanism of triggering and detecting termination.

The GPI implementation follows a similar line of the previous work
with MPI. Whenever a thread finds a solution, it triggers termination
by writing to its peers that it has found a solution. Thus, the time
of the parallel execution is the time taken by this fastest thread.

Other threads must detect termination. This is only possible by
introducing a communication step inside the internal loop of the
Adaptive Search algorithm. This is required since there is no other
way for a GPI instance to react on an remote event (\ie termination)
other than with communication.  In this communication step, a check
for termination is done on a particular memory address that is written
on termination emmission as described above.  The communication step
introduces some overhead that needs to be minimized. Thus the
communication step is only executed every \emph{k} iterations.

\subsection{Propagation of configuration}

The experiments in previous work and with the TDO variant have found
that the simple approach to parallelization, namely, the independent
multiple-walk, proves itself insufficient in obtaining parallel
efficiency on some problems specially when experimenting with a large
number of cores. Moreover, exchanging some simple information such as
the cost leads to no improvement.

Hence, we aim at communicating more and more meaningful information,
introducing cooperation. By cooperation we mean mechanisms that allow
threads to share information about their state and thus benefit from
the collective search. Also, we want to exploit the potential and
benefits of GPI and its programming model (one-sided communication, no
wait for communication, global access to data, threaded model, etc.).

One of the most powerful aspects of Local Search is its
simplicity. And due to this simplicity, it is hard to extract what
could be considered as meaningful information to be shared and
communicated. One logical candidate not yet tried is the whole current
solution or configuration. Because the term \emph{solution} is
sometimes misleading, we refer to the current solution as a
\emph{configuration}. The final solution represents the solution when
the algorithm stops.

The used implementation of the Adaptive Search method deals only with
permutation problems and thus, a configuration is the permutation
vector of the problems' variables.

Similarly to other approaches to the parallelization of local search
methods which introduce cooperation, several important questions
arise, namely:

\noindent
\begin{enumerate}
\item Who does the communication?
\item When to do the communication?
\item How to do the communication?
\item What to communicate?
\end{enumerate}

\noindent
Answering most of these questions requires carrying out actual
experiments since the best and correct answer it is not, in our
opinion, foreseeable.

Our approach, which we call Propagation of Configuration (PoC), aims
at answering these questions and give a better understanding of how
cooperation can help with increasing the scalability of Local Search
in general and the Adaptive Search method in particular.

\subsection*{Who does the communication?}

Answering the question of who does the communication involves deciding
whether a single thread or all threads actually perform
communication. Note that by communication we mean that, in a
distributed setting, messages between nodes are exchanged. In a single
node and given the GPI programming model, we can benefit from the
threaded-model and shared memory. Notwithstanding the best option for
this, it is clear that all threads must benefit from it.

There are potential advantages and disvantages with both options. If
all threads perform communication, any shared resources must be
protected by a mutual exclusion mechanism, which might suffer from
high contention. Moreover, when all threads perform communication a
lot more pressure on the interconnect follows, increasing the parallel
overhead and with possibly a lot of redundant communication happening
(the same configuration being communicated several times). But, on the
other hand, there will be a rapid progress towards the best promising
neighborhood, intensifying the search. Of course, this can be positive
but can also become dangerous since most of threads might get trapped
in a local minimum or poor quality neighborhood. A good trade-off
between intensification and diversification must be achieved.

If a single master thread communicates, the effects are potentially
the opposite: less intensification but also less contention, less
pressure on the interconnect and less redundant work. 

Preliminary tests made clear that the best option is the one with a
single communicating thread since it reduces the parallel
overhead. Plus, with GPI, all threads in a single node benefit
immediately from the results obtained by the master thread without any
exchange of messages.

\subsection*{When to do the communication?}

The first possible answer to this question is to follow the same
strategy as with the Termination Detection Only variant: introduce a
communication step and perform communication every \emph{k}
iteration. The value of \emph{k} is fundamental on how well this
option might perform. With a low value (\eg \emph{k} = 10), a strong
intensification of the search is achieved but with the danger that
threads might give up too soon on a promising neighborhood. With a
high value of \emph{k}, we avoid that danger but less intensification
will be achieved since less information will be propagated. 

The other option is to not interrupt the normal flow of the algorithm
for communication, letting the search progress normally and
independently until a local minimum is achieved. Only at this point
the configuration is propagated and possibly used. One danger however
is if threads don't hit local minima that often, the propagation of
configuration won't progress and some threads might never see an
up-to-date configuration, achieving less intensification. A solution
to this problem is to still have communication every \emph{k}
iteration, where threads simply keep the communication progressing but
only use the propagated information when they are in trouble \ie hit a
local minimum. However, this option increases the overhead by adding
the extra communication step in some iterations.

In principle the second option might seem more promising as no
disturbance is caused when the algorithm is progressing
positively. But the forementioned danger that the propagation of
configurations won't progress can have the consequence that there
won't be a benefit from the communication scheme when compared to the
simple TDO variant. We performed some tests on a problem with low
number of local minima (Magic Square) and in fact, this is what
happens.

Based on this reasoning, our chosen option to when to communicate is
to have a communication step. Moreover, we still need to detect
termination thus a communication step must be present, even if with a
much lower influence in terms of overhead.  Our PoC variant combines
termination detection and the propagation of configurations in a
single step that happens every \emph{k} iterations and we focus on
finding an optimal value for \emph{k}.

\subsection*{How to do the communication?}

With this question, we consider a single alternative. Since we aim at
large scale executions, we need an efficient approach.  Communication
is done in a tree-based topology, in which each node only communicates
with its parent and children (if any). Currently, a binary tree is
used but this can be parametrized at initialization.  At each
communication step, the propagation of the configuration is done
either up (to parent) or down (to the children) the tree. This only
happens if a configuration was propagated from the children (in case
of the up direction) or from the parent (down direction). The
propagation of the communication behaves then like a wave, up and down
the tree, with possibly different configurations being propagated at
different points of the tree and contributing to some diversification.

Communication is performed by using GPI one-sided primitives. A thread
posts a write operation and returns immediately to work. The
configuration to be propagated will be directly written to the memory
of the remote node asynchronously, without any acknowledgement of it
and overlapped with the algorithm's computation. The remote node on
the other hand, on its communication step, checks if a valid
configuration was written to its memory, decides how to act on it and
propagates its decision further.

We consider this single alternative since it gives us a good balance
between intensification and diversification and because having a
tree-based topology provides an efficient pattern to achieve
communication scalability. The final objective is to have a
communication step with low overhead and here GPI provides us with
mechanisms to do so.

\subsection*{What to communicate?}

The Adaptive Search method (as many other Local Search methods) is
very simple and includes very few elements that can be communicated.

The proposed option already mentioned before, is to communicate a full
configuration. To this, we only add the cost of the configuration as
it is the metric to evaluate the configuration. Plus, computing the
cost everytime we communicate a configuration is a source of extra
overhead specially if a problem has a large number of variables.

Still, the question remains of which configuration to communicate. In
our design the best configuration \ie the configuration with better
cost is communicated. At a communication step, a thread decides to
propagate its own current configuration or the propagated
configuration(s). 

Communicating configurations can be of advantage because it includes
implicitly more information about the state of the search since it, in
a sense, provides a semi-exact positioning within the whole search
space. As the best configurations are being propagated, other threads
that are currently on poorer neighborhoods might benefit from moving
to the best ones. With the stochastic behavior of Adaptive Search and
enough diversification, the whole search procedure can be performed on
the best neighborhoods and possibly, converge faster to good
solutions.

\section{Experimental results}
\label{sec:exp_res}
In this section we present the obtained results using different
problems. 

\begin{itemize}
\item \textbf{all-interval}: the All Interval Series problem (prob007
  in CSPLib~\cite{CSPLIB}),
\item \textbf{costas-array}: the Costas Array problem,
\item \textbf{magic-square}: the Magic Square problem (prob019 in CSPLib).
\end{itemize}

\noindent
The experiments were conducted on a cluster system where each node
includes a dual Intel Xeon 5148LV (``Woodcrest'') (\ie 4 CPUs per
node) with 8 GB of RAM. The full system is composed of 620 cores
connected with Infiniband (DDR).  Since we aim at large scale, we
performed our experiments on the system using up to 256 cores on some
problems and 512 cores on others. This difference is due to the fact
that the system is largely used and is hard to get access to the full
system.

Note that Adaptive Search, as many other Local Search methods, has a
stochastic behavior to achieve diversity on the search. To benchmark
such behavior, several executions must be done and averaged. In our
experiments we ran each problem 100 times in order to obtain
meaningful results.

We compare both GPI variants (TDO and PoC) with the MPI implementation
from previous work, which serves as our basis for comparison.

\begin{figure}[h!]
\centerline{\includegraphics[scale=0.6]{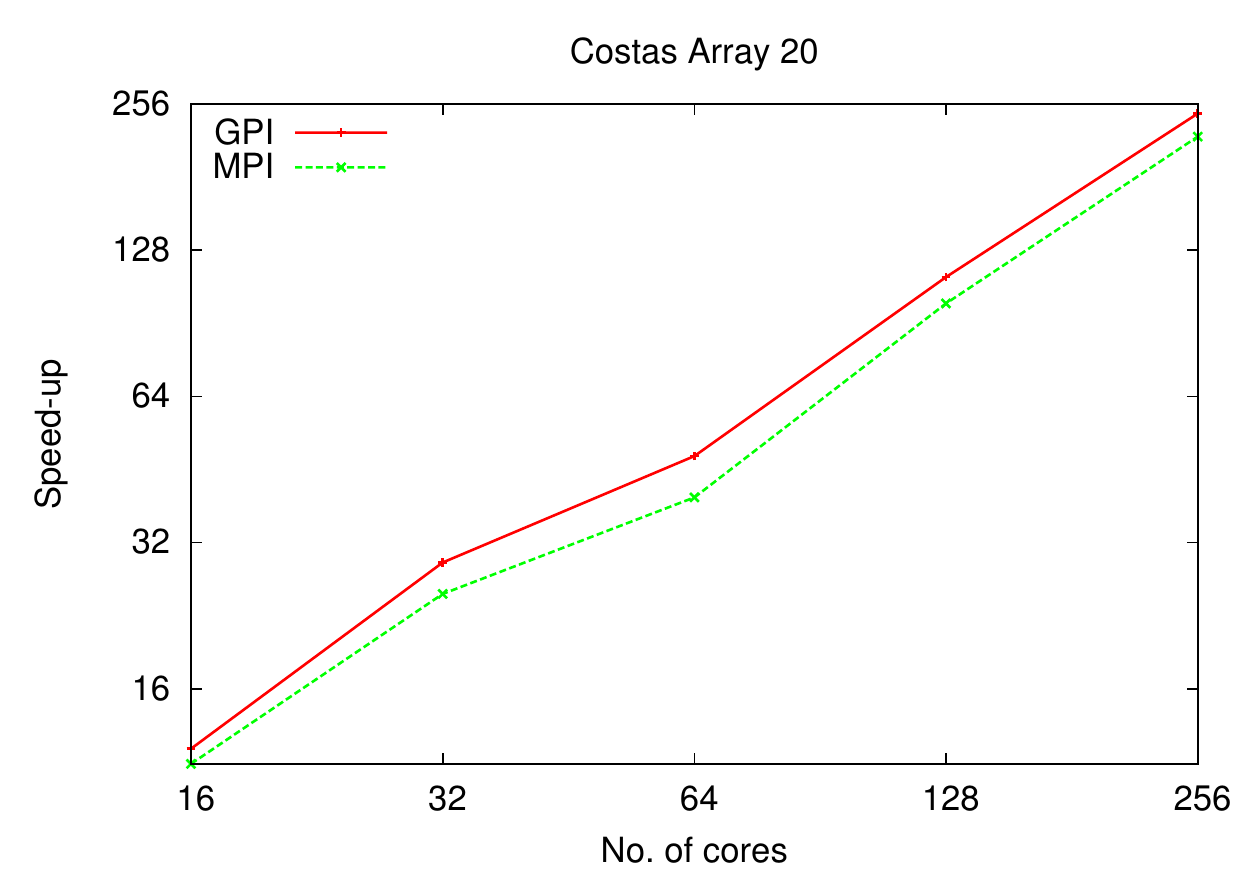}}
\caption{Costas Array (n=20) on 256 cores (64 nodes)}\label{fig:costas20}
\end{figure}

Fig.~\ref{fig:costas20} depicts the obtained results for the Costas
Array problem (CAP) with \emph{n}=20.

As already observed in previous work~\cite{Caniou:2012}, the CAP shows
an almost optimal scalability using an independent multiple-walk with
no cooperation. We can observe that our implementation obtains
similar, although slight better, results. This is the expected result
since both approaches (TDO and MPI) are equivalent and a confirmation
that our implementation performs as expected.

Although we aspired at obtaining even better results with the PoC
variant (possibly super linear) for this problem, our experiments
showed that this variant performs much worse than the simple TDO
variant and thus we only present the speedup obtained with GPI using
the TDO variant.

\begin{figure}[h!]
\centerline{\includegraphics[scale=0.6]{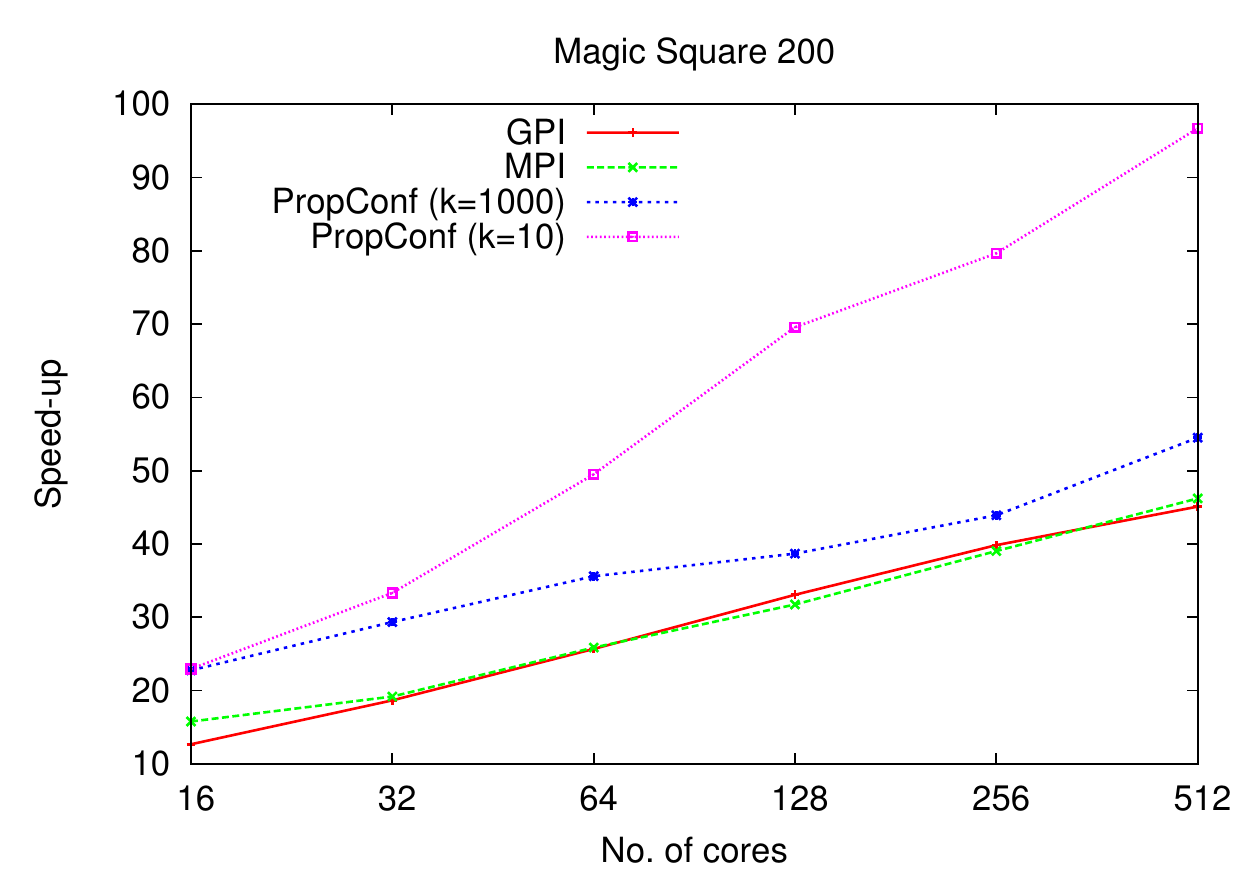}}
\caption{Magic Square 200 on 512 cores (128 nodes)}\label{fig:ms200}
\end{figure}

The Fig.~\ref{fig:ms200} depicts the obtained results for the Magic
Square problem up to 512 cores.

For this problem we present the speedup obtained with the TDO and PoC
variants and compare it with the MPI version. The GPI TDO variant
presents again, as expected, results similar to the MPI version.

The Magic Squares benchmark is one of the problems that results in
disappointing scalability when using the simple independent multiple-walk
and therefore a major target for improvement with more sophisticated
approaches. Indeed, for this problem, our PoC variant improves the
performance and scales better as we increase the number of cores used.

We wanted to answer the question of when to do communication: as we
mentioned, in our preliminary experiments it turned out that the best
approach is to have a communication step every \emph{k} iterations
where the value of \emph{k} is decisive. Surprinsingly, for this
benchmark, a lower value of \emph{k} (\emph{k}=10 in contrast to
\emph{k}=1000) improves scalability by a factor of 2, achieving a
speedup of 97 with 512 cores. Still a low parallel efficiency but a
large improvement over the other options and variants.

\begin{figure}[h!]
\centerline{\includegraphics[scale=0.6]{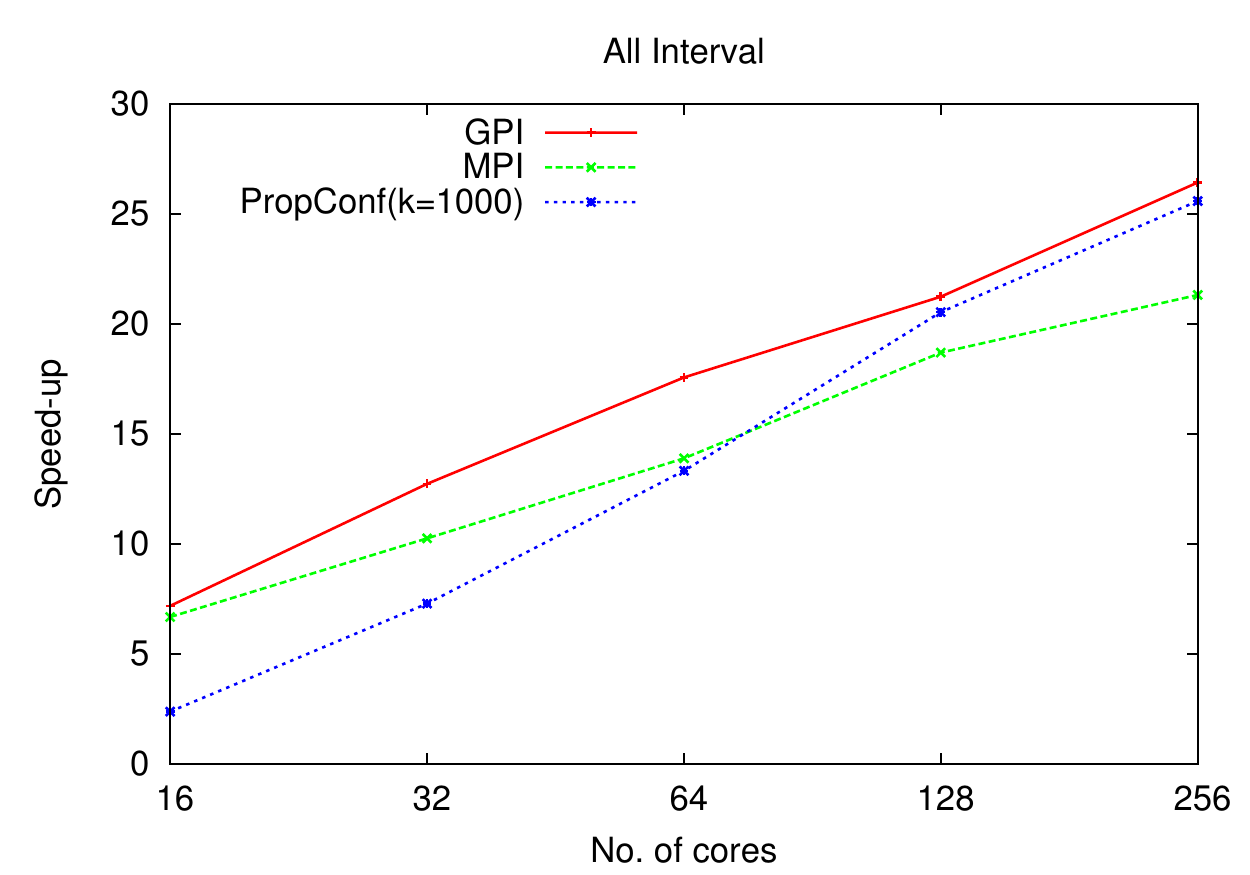}}
\caption{All Interval 400 on 256 cores (64 nodes)}\label{fig:ai400}
\end{figure}

The obtained results for the last problem, the All Interval series
(\emph{n}=400), is shown in Fig.~\ref{fig:ai400}.

The All Interval Series benchmark is also one of the problems where
good scalability was hard to reach when using a large number of cores.
In Fig.~\ref{fig:ai400} it is possible to observe this fact, where
both the MPI and GPI TDO versions reach a modest speedup factor of 20
and 25, respectively (with 256 cores). Our PoC variant however,
performs much worse than the TDO variant at a low number of cores but
it improves as we increase the number of cores, hinting that this
variant can be of advantage if we increase the number of cores and the
problem size.  In Fig.~\ref{fig:ai400} we only depict the obtained
results for the PoC variant with \emph{k} = 1000 since, for this
benchmark, it is the best value. In contrast to the Magic Squares
benchmark, a lower value of \emph{k} results in a much worse
performance.

\section{Discussion}
\label{sec:discussion}

The experimental results presented large differences in how the
different problems benefit from parallelism and the implemented
variants. One of our main objectives is to investigate and understand
why this happens.

In order to be able to draw some conclusions on our experiments, it is
important to characterize the chosen problems from different
perspectives. We characterize the problems using different information
such as the number of iterations and local minima. This
characterization will give us a basis to better understand the
problems at hand and possibly explain our results.

The Table~\ref{tab:prob_charac} presents the obtained values for acquired
information when running some instances of the previously presented
problems. This information is the following:

\begin{description}
\item[Problem] The problem instance.
\item[Iterations] The number of iterations required to find a solution.
\item[Local Minima] The number of local minima found.
\item[Resets] The number of partial resets performed (not full restart).
\item[Same var / Iteration] The number of times that existed more than
  one candidate variable (highest error value) to be selected.
\end{description}

\noindent
This information allows us to better understand how does the Adaptive
Search algorithm progress towards a solution, the neighborhood
structure and extract further information (\eg number of local minima
\emph{per} iteration).

\begin{table}[h]
\renewcommand{\arraystretch}{1.5}
\centering
\begin{tabular}{|c|c|c|c|c|}
\hline
Problem	 & Iterations & Local Minima & Resets &	Same var/Iteration \\ \hline
Magic Square 200 & 413900.505 & 25864.75 & 3.01 & 23.36 \\ \hline
Costas 18 & 389932.263 & 204024.89 & 204024.89 & 1.00   \\ \hline
Costas 19 & 3364807.772	& 1714299.50 & 1714299.50 & 0.99 \\ \hline
All Interval 200 & 11229.220 & 495.27 & 495.27 & 5.97 \\ \hline
All Interval 400 & 41122.406 & 1422.15 & 1422.15 & 9.19 \\ \hline
\end{tabular}
\caption{Information collected for different problems instances.}
\label{tab:prob_charac}
\end{table}

From Table~\ref{tab:prob_charac} we can see that the different
problems exhibit a quite different behavior. The Magic Square problem
performs a low number of partial resets when compared to the total
number of iterations or to the number of identified local minima. On
the other hand, it is the problem where the number of candidate
variables per iterations (Same var/Iteration) is high, meaning that at
each iteration there are several possible moves towards the next
configuration.

The Costas Array problem exhibits a completely different behavior. In
this problem, the number of local minima identified is very large
(almost every second iteration finds a local minimum) and the number
of partial resets is very high, coincident with the number of local
minima \ie at each local minimum found, a partial reset is
performed. Also the number of possible moves at each iteration is
close to 1.

The All Interval problem is yet another kind problem. Here, the
number of resets is as with the CAP equal to the number of local
minima but these happen much less often. The number of possible
variable choices or moves is higher than 1, meaning that some
diversification could be achieved. 

If we relate this characterization of problems with the obtained
experimental results, some conclusions can be conjectured in order to
better understand the parallelization of such algorithm or, more
concretely, how much can it benefit from a communication scheme
such as the one we designed.

We argue that one critical aspect is the neighborhood of a
configuration or the set of possible moves, which define transitions
between configurations. Since we are propagating configurations we can
look at our problems at hand according to this aspect. If a problem
has a dense neighborhood or, in other words, the set of possible moves
at each transition is (much) larger than one, each of these moves can be
explored in parallel. Thus, when a promising configuration is
propagated and several moves are possible and explored in parallel,
the probability that one of these moves leads to a faster path towards
an optimal solution increases.

Another important aspect is the number of local minima and resets and
how both relate. A problem that finds a large number of local minima
before encountering an optimal solution benefits less from processing
a configuration which seems promising. This configuration is
heuristically promising but in reality this information is less
meaningful than it should. Similarly, a problem with a high number of
partial resets suffers from the same problem.

Looking back at our experimental results with the different problems,
we can better understand a) the difference in scalability and b) the
improvement factor brought by the PoC variant to some problems.

In the Magic Square problem, each configuration has a dense
neighborhood and benefits from the parallel exploration of different
moves. Thus, the PoC variant improves the performance and scalability
of the algorithm. When a working thread adopts a propagated
configuration, it will define its own path from that configuration and
differently from one other thread that receives that same promising
configuration. Moreover, this problem has a low number of local minima
and resets meaning that paths from one (initial) configuration towards
an optimal solution are a series of transitions from neighbor
configurations.

The Costas Array Problem exhibits optimal scalability with the
independent multiple-walk MPI version or with our TDO variant and this
is already \emph{per se} satisfactory. On the other hand, it performs
worse with the PoC variant: propagating a configuration is only a
source of parallel overhead and will limit the search allowing less
diversification. A propagated configuration will allow, on average, a
single move and two threads taking the same configuration results in
redundant work which is also probably unfruitful since the CAP is one
of the problems with a high number of local minima and reset. This
also explains the good scalability using the TDO variant, where
increasing the number of cores allows covering more of the total
search space together with the fact that solutions for this problem
are well spread over it.

Finally, the All Interval Series problem shows a mixed
behavior. Similarly to the CAP, the larger number of local minima
found and same number of resets point to the same problem. There is
less benefit from taking a propagated configuration since its
meaningfulness is low. The PoC variant only introduces unnecessary
overhead and this could explain the much worse performance at a lower
number of cores. On the other hand, and similarly to the Magic Square
benchmark, there is more than one possible move, on average \ie some
diversification can be achieved. With a large enough number of cores,
the parallel overhead can be amortized by the gain obtained with this
diversification. This could be the reason for the steeper curve for
the PoC variant on Fig.~\ref{fig:ai400}. Of course, with further
experiments we will be able to understand this better.

In summary, problems where configurations have a denser neighborhood
benefit from a cooperation scheme such as the PoC variant where the
full configuration is communicated. Contrarily, problems that follow a
trajectory with a single move possible won't benefit from a
communication scheme that propagates the best current configuration.
Also, if a large number of local minima is found and partial resets
are required in the same number, the expectation for improvement in
performance is zero.

\section{Conclusion}
\label{sec:conclusion}

In this paper we presented our work on the parallel implementation of
the Adaptive Search method using a different programming model. GPI is
an API designed for high-performance and scalable parallel
applications.  We aimed at investigating and understanding the
behavior of Adaptive Search in a parallel setting, focusing on
different problems particularly those that, in previous work, showed
scalability problems when targeting a large number of cores.  GPI and
its programming model allowed us to design a new communication and
parallelization scheme which in our experimental evalution allowed a
gain of a factor of 2 in terms of speedup for some problems. More
importantly, it provided deeper insight and understanding on the
parallelization of Local Search methods given different problems with
disparate characteristics such as the neighborhood of a configuration,
the number of local minima and partial resets.  

In the future, we intend to examine our design and conclusions with
other larger problems and experiment with complexer parallelization
schemes. One possible direction is instead of using promising
information (configurations, cost, statistics) directly, act on the
complement of it, avoiding redundant work and cover as much as
possible from the search space since this is the main source of
parallelism.

One of our potential final goals is the design of a new Local Search
algorithm more amenable to parallelization that builds upon these
experiences.

\textbf{Acknowledgements} This work was partly supported by
Funda\c{c}\~ao para a Ci\^encia e Tecnologia under grant
PTDC/EIA-EIA/100897/2008 (HORUS).

%
%
\bibliography{paper}
\bibliographystyle{plain}

\end{document}